\documentclass[journal=jacsat,manuscript=article]{achemso}

\usepackage{chemformula} % Formula subscripts using \ch{}
\usepackage[T1]{fontenc} % Use modern font encodings
\usepackage[version=3]{mhchem} % Formula subscripts using \ce{}
\usepackage{multirow}
\usepackage{xcolor}
\usepackage{epsfig}
\usepackage{graphicx}
\author{Chen Shen}
\affiliation[Technical University Darmstadt]
{Institute of Materials Science, Technical University Darmstadt, Alarich-Weiss-Strasse 2, 64287 Darmstadt, Germany}%\altaffiliation{Department of Materials Science, TU Darmstadt, Germany}
\author{Qiang Gao}
\affiliation[Technical University Darmstadt]
{Institute of Materials Science, Technical University Darmstadt, Alarich-Weiss-Strasse 2, 64287 Darmstadt, Germany}%\altaffiliation{Department of Materials Science, TU Darmstadt, Germany}
\author{Nuno M. Fortunato}
\affiliation[Technical University Darmstadt]
{Institute of Materials Science, Technical University Darmstadt, Alarich-Weiss-Strasse 2, 64287 Darmstadt, Germany}\author{Harish K. Singh}
\affiliation[Technical University Darmstadt]
{Institute of Materials Science, Technical University Darmstadt, Alarich-Weiss-Strasse 2, 64287 Darmstadt, Germany}\author{Ingo Opahle}
\affiliation[Technical University Darmstadt]
{Institute of Materials Science, Technical University Darmstadt, Alarich-Weiss-Strasse 2, 64287 Darmstadt, Germany}
\author{Oliver Gutfleisch}
\affiliation[Technical University Darmstadt]
{Institute of Materials Science, Technical University Darmstadt, Alarich-Weiss-Strasse 2, 64287 Darmstadt, Germany}

\author{Hongbin Zhang}
\email{hzhang@tmm.tu-darmstadt.de}
\affiliation[Technical University Darmstadt]
{Institute of Materials Science, Technical University Darmstadt, Alarich-Weiss-Strasse 2, 64287 Darmstadt, Germany}

\title[An \textsf{achemso} demo]
  {Designing of Magnetic MAB Phases for Energy Applications}


\begin{document}

\begin{tocentry}

\centering 
\includegraphics[width=9cm, height=3.5cm]{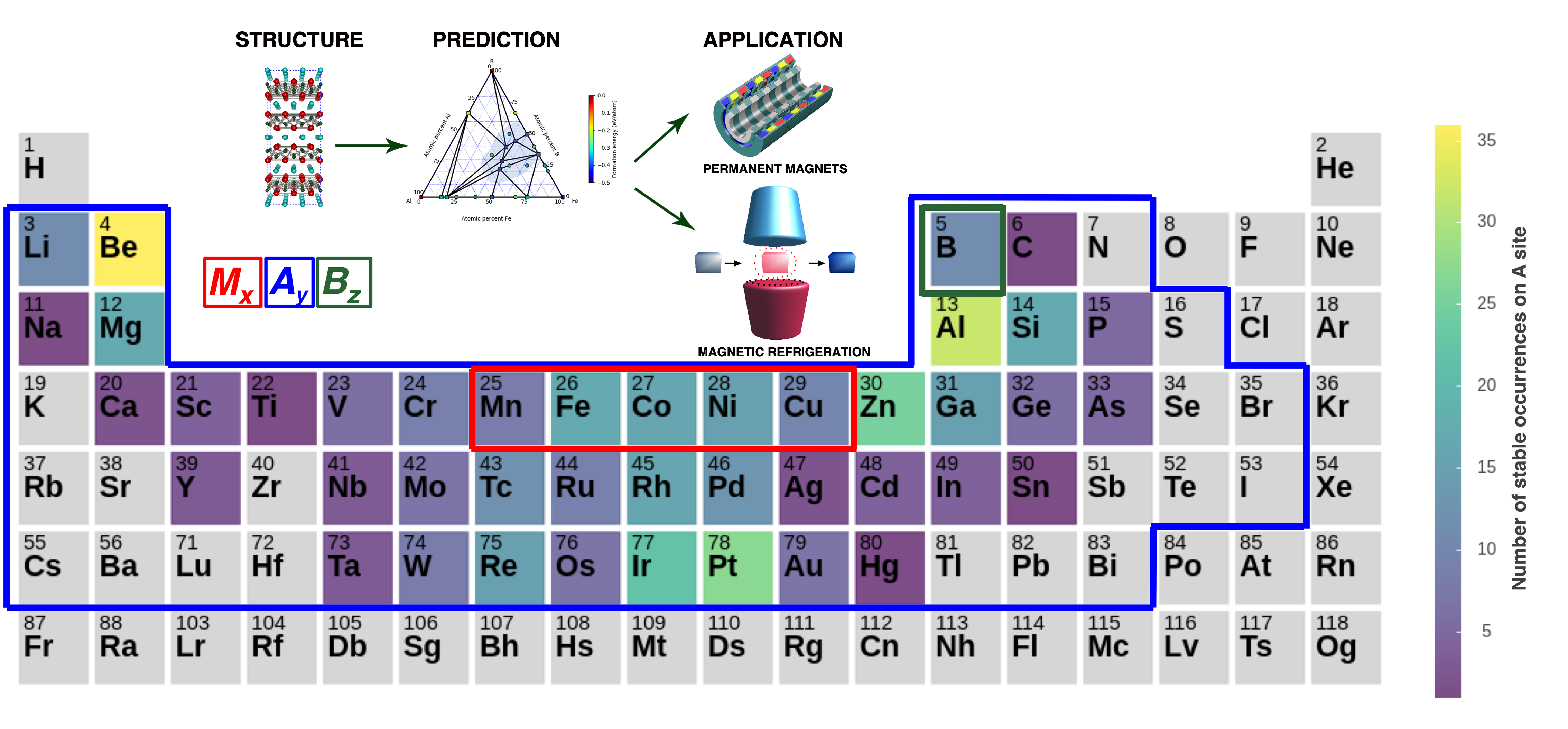}

\end{tocentry}

\begin{abstract}
Based on high-throughput density functional theory calculations, we performed screening for stable magnetic MAB compounds and predicted potential strong magnets for permanent magnet and magnetocaloric applications.
The thermodynamical, mechanical, and dynamical stabilities are systematically evaluated,
resulting in 21 unreported compounds on the convex hull, and 434 materials being metastable considering convex hull tolerance to be 100 meV/atom.
Analysis based on the Hume-Rothery rules revealed that the valence electron concentration and size factor difference are of significant importance in determining the stability, with good correspondence with the local atomic bonding.
We found 71 compounds with the absolute value of magneto-crystalline anisotropy energy above 1.0 MJ/m$^3$ and 23 compounds with a uniaxial anisotropy greater than 0.4 MJ/m$^3$, which are potential gap magnets. 
Based on the magnetic deformation proxy, 99 compounds were identified as potential materials with interesting magnetocaloric performance. 
\end{abstract}

\section{Introduction}
The modern industrial and societal demands for advanced functional magnetic materials are growing faster as we are witnessing the global expansion of hybrid-electric vehicles, robotics, wind turbines, and automation, leading to a strong incentive on the green energy revolution~\cite{gutfleisch2011magnetic,skokov2018heavy}.
Particularly, the efficient harvesting of renewable energy (such as wind energy) and endeavor to reduce the greenhouse effect (mainly through the development of e-mobility and magnetic refrigeration) have intensified the impetus to design resource-efficient magnetic materials with optimal performance, such as permanent magnets and magnetocaloric materials. 
For instance, one interesting question is to identify the so-called gap magnets~\cite{coey2012permanent}, {\it i.e.}, permanent magnets with their energy density (BH)max~\cite{gutfleisch2000controlling} lying between the widely applied AlNiCo and Ferrites~\cite{sugimoto2011current} and the high-performance Sm-Co~\cite{strnat1991rare} and Nd-Fe-B-based~\cite{herbst1991neodymium} permanent magnets.
Potential candidates can be characterized using the dimensionless figure of merit $\kappa = \sqrt{K_1/(\mu_0M_S^2)}$~\cite{coey2014new}, providing an effective descriptor for high-throughput screening.
Moreover, following the discovery of Gd$_5$Si$_2$Ge$_2$~\cite{pecharsky1997giant} and LaFeSi$_{13}$~\cite{gutfleisch2005large} with giant magnetocaloric effect (MCE) around room temperature, magnetic refrigeration technology is assumed to be capable of competing
with and hopefully surpassing conventional refrigeration in terms of energy efficiency, environmentally friendly and ecological impact in the near future~\cite{gutfleisch2016mastering,scheibel2018hysteresis, gottschall2019making}.
However, most permanent magnets and potential magnetocaloric materials with high performance are based on the intermetallic compounds containing rare-earths (RE), which are resource critical~\cite{sander20172017}.
Therefore, rare-earth-free permanent magnets and MCE materials with enhanced efficiency over a broad temperature range and useful secondary properties, such as mechanical stability, corrosion resistance, shapeability, sustainability, and recyclability, are still desirable~\cite{mohapatra2018rare,gutfleisch2011magnetic,kuz2014towards}.

The MAB phases with nanolaminated crystal structures exhibit intriguing magnetic properties and mechanical deformation behavior, which have attracted considerable attention recently~\cite{kota2019progress}.
Such materials are ternary borides comprising stacked M-B layers (M = transition metal, B = Boron) interleaved by monolayers of A atoms. 
In this regard, the crystal structures are quite similar to those of the well-known MAX phases M$_{n+1}$AX$_n$ (X = C and N, A denotes a main group element), 
which host a unique combination of metallic and ceramic properties~\cite{barsoum2000mn+}. 
The novel magnetic nanolaminates recently discovered in the MAX phases~\cite{ingason2016magnetic}, 
are also expected in the MAB phases.
Moreover, Fe$_2$AlB$_2$ was found to be a promising magnetocaloric material exhibiting an interesting MCE~\cite{tan2013magnetocaloric}, 
with the ordering temperature around 300 K confirmed by experimental~\cite{cedervall2016magnetic,barua2018anisotropic} and theoretical studies~\cite{chai2015investigation,kadas2017alm2b2,el2019magnetocaloric}. 
Ke {\it et al.}~\cite{ke2017electronic} investigated the intrinsic properties of Fe$_2$AlB$_2$, and found a MAE as large as -1.34 MJ/cm$^3$, in good agreement with the experiments~\cite{barua2018anisotropic}.
Recently, Cr$_4$AlB$_4$ with a novel structure of MAB phase has been synthesized consistent with the theoretical calculations~\cite{zhang2019crystal}.
Khazaei {\it et al.}~\cite{khazaei2019novel} carried out high-throughput (HTP) calculations on Al-containing non-magnetic MAB phases and predicted 9 stable compounds. 
More recently, Miao {\it et al.}~\cite{doi:10.1021/acs.chemmater.0c02139} reported another HTP screening for Ti-A-B, Zr-A-B, and Hf-A-B and predicted 7 thermodynamically stable compounds.
Therefore, an interesting question is whether there exist more stable MAB compounds beyond the above-mentioned cases and whether are they good candidates as potential functional magnetic materials.

\begin{figure}
    \centering
    \includegraphics[width=15cm]{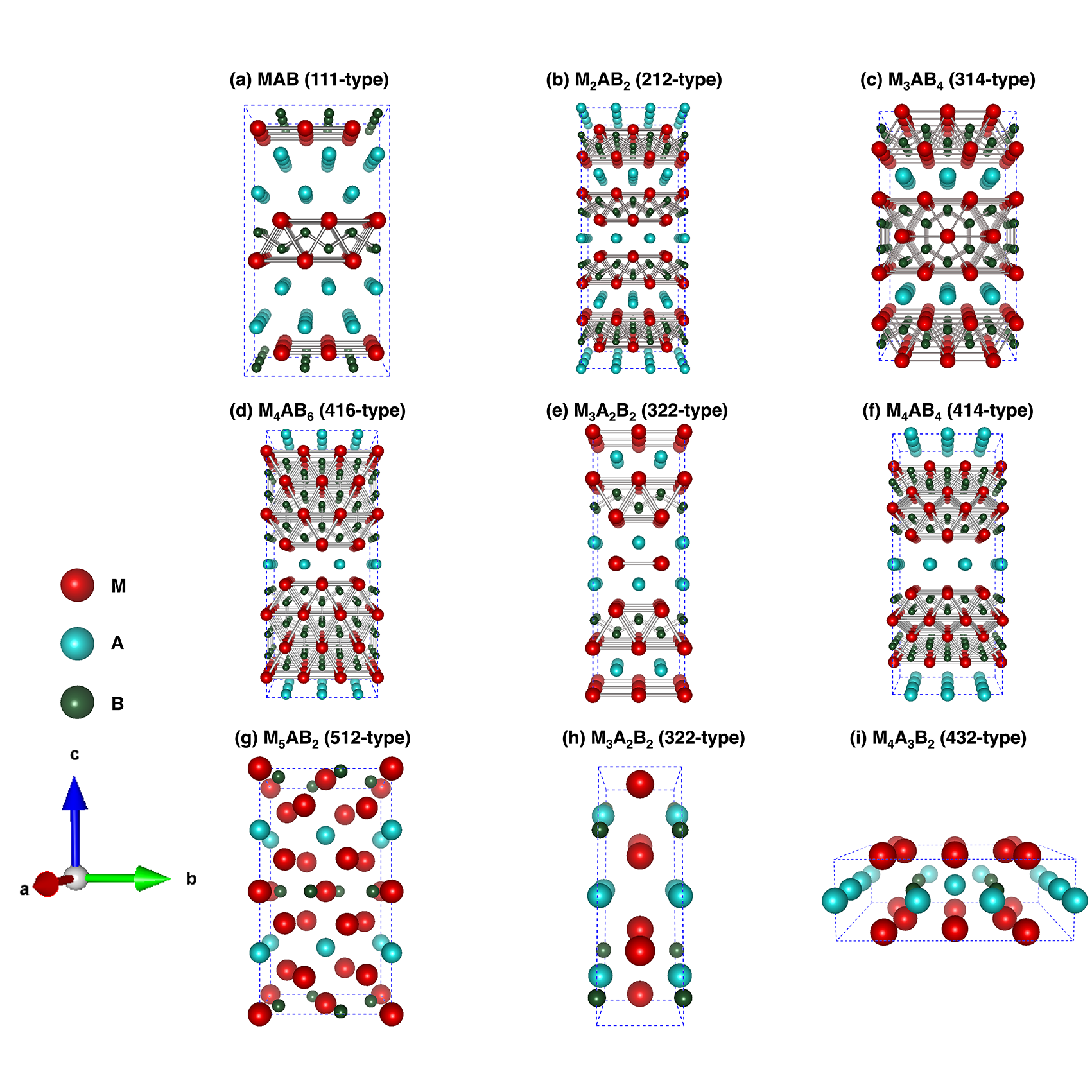}
    \caption{Crystal structures of considered MAB phases(a-f) and non-MAB phases(g-h): (a)222-type [Cmcm], (b)212-type [\textit{Cmcm}], (c)314-type [\textit{Pmmm}], (d)416-type [\textit{Cmmm}], (e) 322-type [\textit{Cmcm}], (f) 414-type [\textit{Immm}], (g) 512-type [\textit{I4/mcm}], (h) 322-type [\textit{P2/m}] and (i) 432-type [\textit{P4/mmm}].}
    \label{all-structure}
\end{figure}

In this work, based on HTP density functional theory (DFT) calculations, we systematically studied the stabilities and the magnetic properties of the MAB compounds to identify possible candidates for permanent magnets and magnetocaloric materials.
Six experimentally synthesized MAB phases and three non-MAB phases (as competitive phases) are considered (Fig. \ref{all-structure}), 
including MAB~\cite{jeitschko1966kristallstruktur} (space group \textit{Cmcm}), M$_2$AB$_2$~\cite{jeitschko1969crystal} (space group \textit{Cmcm}), M$_3$A$_2$B$_2$~\cite{jung1986kristallstrukturen} (space group \textit{Cmcm}), M$_3$AB$_4$~\cite{chaban1973ternary} (space group \textit{Immm}), M$_4$AB$_4$~\cite{zhang2019crystal} (space group \textit{Immm}) and M$_4$AB$_6$~\cite{ade2015ternary} (space group \textit{Cmcm}); non-MAB phases are M$_5$AB$_2$~\cite{haggstrom1975mossbauer}(space group \textit{I4/mcm}), M$_3$A$_2$B$_2$~\cite{hirt2018synthesis} (space group \textit{P2/m}) and M$_4$A$_3$B$_2$~\cite{hirt2018synthesis} (space group \textit{P4/mmm}). 
Three non-MAB phases are considered as competitive phases in order to make the prediction of MAB compounds more reliable.
Such compounds are flexible in the chemical compositions and have tunable magnetic properties. 
For example, Fe$_5$SiB$_2$ has a T$_C$ higher than 760 K, a M$_S$ larger than 1 MA/m, and a MAE more than 0.30 MJ/m$^3$ at room temperature~\cite{cedervall2016magnetostructural,mcguire2015magnetic,werwinski2016magnetic,lamichhane2016study,hedlund2017magnetic}.
After validating all the experimentally known phases, we identified stable and metastable ternary borides based on the systematic evaluation of the thermodynamical, mechanical, and dynamical stabilities.
Taking the M$_2$AB$_2$-type compounds as an example, we investigated the effect of magnetic ordering on the thermodynamic stability, followed by a comprehensive analysis of the stability trend following the Hume-Rothery rules and local atomic bonding.
The MAE and magnetic deformation proxy are evaluated explicitly, which help to screen for potential permanent magnets and magnetocaloric materials. 
Our work expands the materials library of rare-earth free permanent magnets and magnetic refrigeration, and thus provides valuable guidance to further theoretical and experimental studies to design advanced magnetic materials in transition metal-based ternary borides for energy applications~\cite{kuz2014towards,gottschall2019making}.

\section{Computational details}
The DFT calculations are performed using an in-house developed HTP environment~\cite{opahle2012high,opahle2013high} to determine the thermodynamical stability for the above mentioned six MAB and three non-MAB phases, as demonstrated in recent studies~\cite{gao2020designing, PhysRevMaterials.3.053803, singh2018high}. 
It is noted that the non-MAB phases are regarded as competitive phases for the MAB phase to obtain the reliable convex hull, which is also applied in designing MAX phases by considering antiperovskites as a competitive phase~\cite{PhysRevMaterials.3.053803}.  
Thermodynamical stability is evaluated by considering the formation energy (E$_f$) and the distance to the convex hull with respect to all the relevant competing phases available in the OQMD database~\cite{saal2013materials}. 
All the calculations are carried out using the Vienna ab initio simulation package (VASP) code~\cite{kresse1996efficient,kresse1999ultrasoft}.
The MAE of the predicted stable phases is obtained using the full-potential local-orbital (FPLO)~\cite{koepernik1999full} code in the force theorem regime,
and the recently proposed magnetic deformation proxy~\cite{bocarsly2017simple} is used to evaluate the MCE. 
More details of the computational processes can be found in the Supplementary.

\section{Results and discussion}
\subsection{Stabilities of phases}
\subsubsection{Thermodynamical stability}
The thermodynamical stability of the MAB and non-MAB phases (shown in Fig.~\ref{all-structure}) are obtained based on the formation energy $\Delta E_f$ and distance to the convex hull $\Delta E_h$, where $\Delta E_f < 0$ and $\Delta E_h = 0$ are required for the stable phases.
In general, $\Delta E_f<0$ ensures that the target compounds are energetically stable against decomposing into the constituent elements following the reaction $M_xA_yB_z \rightarrow xM + yA + zB$,
whereas $\Delta E_h=0$ denotes the stability upon the decomposition into any binary and ternary phases according to the reaction $\Delta E_h = E_{tot}(\text{predicted phase}) - E_{tot}(\text{competing phases})$.
In our calculations, the competing phases include all the relevant compounds found in the OQMD database~\cite{saal2013materials,kirklin2015open}.
As summarized in Table.~\ref{tab:stable}, there are 21 compounds satisfying the thermodynamic stability criteria, 17 of them are with one of the MAB structures.
According to the literature, 15 ternary borides with one of the considered structures have been experimentally synthesized, as listed in Table.~\ref{tab:stable}.
All such compounds exhibit $\Delta E_f < 0$ and $\Delta E_h < 80 $ meV/atom (due to numerical errors in DFT), validating our methodology and hence the validity of the newly predicted phases.
The resulting lattice parameters are in good agreement with the existing measurements and other theoretical calculations, as listed in Table.~\ref{tab:stable}.
A special case is Co$_5$PB$_2$, where the lattice constants are underestimated (overestimated) along [100] ([001]) directions. 
This is also observed in previous DFT calculations,~\cite{werwinski2018magnetocrystalline} which may be driven by the
missing spin fluctuations as confirmed in (Fe$_{1-x}$Co$_x$)$_2$B~\cite{zhuravlev2015spin}.

Furthermore, not all the compounds are magnetic, {\it e.g.}, with finite magnetization larger than 0.05 $\mu_B$ per magnetic atom (Table.~\ref{tab:stable}).
It is observed that the nonmagnetic compounds occur mostly for the Cr-, Mn-, and Ni-based cases, whereas Fe$_4$Al$_3$B$_2$ and Co$_4$Be$_3$B$_2$ are nonmagnetic as well.
This can be understood based on the Stoner criteria, where $I\nu(E_F)>1$ indicate possible itinerant magnetic ordering, with $I$ being the Stoner parameter and $\nu(E_F)$ the density of states (DOS) at the Fermi energy $E_F$ of the nonmagnetic state.
For instance, the Stoner parameters of magnetic atoms range between 0.6 and 0.75 from Cr to Ni,~\cite{oppeneer2001handbook}, thus those compounds with marginal $\nu(E_F)$ smaller than 1.4 states/eV per magnetic atom end up as nonmagnetic (Fig. S3) because 
$I\nu(E_F) < 1$.
Moreover, the predicted results agree well with previous experimental and theoretical reports, {\it e.g.}, Fe$_5$PB$_2$ with average magnetization 1.71\,$\mu_B/\text{Fe}$~\cite{lamichhane2016study} and Cr$_4$AlB$_{4/6}$ being nonmagnetic.~\cite{ade2015ternary,zhang2019crystal} 

\begin{table}
\small
\caption{List of MAB and non-MAB phases that we found stable based on relative stability analysis. The present considered phases experimentally synthesized are indicated by asterisks (*). Lattice parameters (\AA), formation energy (eV/atom), distance to the convex hull (ev/atom), competing phases, magnetism and magnetic moment ($\mu B/Mag$) in considered phases are shown.}
\label{tab:stable}
\resizebox{\textwidth}{100mm}{
\begin{tabular}{lcccccccccc}
\hline
\hline
\multirow{2}{*}{MAB phase} & \multirow{2}{*}{Space group} & \multirow{2}{*}{a} & \multirow{2}{*}{b} & \multirow{2}{*}{c}  & \multirow{2}{*}{$\Delta E_f$} & \multirow{2}{*}{$\Delta E_h$} & \multirow{2}{*}{Most competing phases} &\multirow{2}{*}{Magnetism}& \multirow{2}{*}{M} \\ \\
\hline
FeBeB & 63 &2.648  & 12.164 & 2.925 &  -0.326 & 0 &FeB, Be$_2$Fe, B  &FM& 0.422 \\ \hline
MnBeB & 63 &2.811  & 12.252 & 2.809 &   -0.378 & 0 &MnB, Be    &NM& 0.002 \\ \hline

Fe$_2$AlB$_2$ $^*$ & 65 &2.916  &11.019  &  2.851  & -0.401 & 0 & FeAl$_6$, AlB$_2$, FeB &FM& 1.330 \\
Ref. Exp.~\cite{tan2013magnetocaloric}        &  &2.928  &11.033  &  2.868  &   &  &  &  &\\
Ref. Cal.~\cite{ke2017electronic}       &  &2.915  &11.017  &  2.851  &   &  &  &  &\\ \hline

Fe$_2$BeB$_2$ & 65 &2.904  &9.947  &  2.749  & -0.344 & 0 & Be$_2$Fe, B, FeB &AFM& 0.760 \\ \hline

Cr$_2$AlB$_2$ $^*$ & 65 & 2.923 & 11.051 & 2.932   & -0.466 & 0 &Cr$_3$AlB$_4$, Cr$_7$Al$_{45}$, CrB  &NM& 0.010 \\ 
Ref. Exp.~\cite{ade2015ternary}        &  &2.937  &11.051  &  2.968  &   &  &  &    &\\
Ref. Cal.~\cite{ke2017electronic}       &  &2.921  &11.034  &  2.929  &   &  &  &    &\\ \hline

Mn$_2$AlB$_2$ $^*$ & 65 &2.894  & 11.080 & 2.831 &   -0.471 & 0 &Mn$_4$Al$_{11}$, MnB, MnB$_4$  &AFM& 0.765 \\
Ref. Exp.~\cite{kadas2017alm2b2}        &  &2.923  &11.070  &  2.899  &   &  &  &  AFM & \\
Ref. Cal.~\cite{ke2017electronic}       &  &2.887  &11.109  &  2.830  &   &  &  & AFM  & \\ \hline

Mn$_2$BeB$_2$ &  65&2.846  & 9.969 &2.815  &   -0.435 & 0 & MnB, Be &NM&  0.011  \\ \hline

Cr$_3$AlB$_4$ $^*$ & 47 &2.939  & 2.939 & 8.091  & -0.445 & 0 & Cr$_2$AlB$_2$, CrB$_4$, CrB &NM& 0.049  \\
Ref. Exp.~\cite{ade2015ternary}        &  &2.956  &2.978  &  8.054  &   &  &  &  & &  \\
Ref. Cal.~\cite{li2019elastic}       &  &2.938  &2.943  &  8.090  &   &  &  &  &  &\\ \hline

Cr$_4$AlB$_6$ $^*$ & 65 &2.947  & 21.328 & 2.943 &   -0.422 & 0.012 &CrB$_4$, Cr$_3$AlB$_4$, CrB  &NM& 0.003  \\
Ref. Exp.~\cite{ade2015ternary}        &  &2.952  &21.280  &  3.013  &   &  &  &  & & \\
Ref. Cal.~\cite{zhou2017electrical}       &  &2.972  &21.389  &  2.961  &   &  &  &  & & \\ \hline

Fe$_4$AlB$_4$ & 71  & 2.927 &18.565 & 2.870 &   -0.417 &0  &AlFe$_2$B$_2$, FeB  &FM& 1.271  \\ \hline
Fe$_4$BeB$_4$ & 71 & 2.918 &17.513 & 2.821 &   -0.377 & 0 & FeB, Be$_2$Fe, B &FM& 1.017  \\ \hline 
Fe$_4$GaB$_4$ & 71  &2.939  & 18.557 & 2.883 &   -0.343 &0  &FeB, Ga$_3$Fe, B  &FM& 1.288  \\ \hline
Fe$_4$MgB$_4$ & 71 &2.932  & 19.626 & 2.875 &   -0.354 &0  & FeB, Mg &FM& 1.391  \\ \hline

Fe$_4$ZnB$_4$ & 71 & 2.931 & 18.726 &2.872&   -0.348 & 0 & FeB, Zn &FM&  1.326  \\ \hline

Cr$_4$AlB$_4$ $^*$ & 71 &2.920  &18.856  &2.939  &   -0.510 & 0 & AlCr$_2$B$_2$, CrB &NM& 0  \\
Ref. Exp.~\cite{zhang2019crystal}        &  &2.934  &18.891  &  2.973  &   &  &  &  & & \\
Ref. Cal.~\cite{zhang2019crystal}       &  &2.932  &18.912  &  2.957  &   &  &  &  & & \\ \hline

Mn$_4$BeB$_4$ & 71 & 2.899 & 17.591 & 2.878 &   -0.467 & 0 & MnB, Be  &FM& 0.878  \\ \hline

Mn$_4$AlB$_4$ & 71 & 2.929 & 18.591 & 2.889 &   -0.499 & 0 & MnB, Mn$_2$AlB$_2$ &FM& 1.014  \\ \hline

Mn$_4$IrB$_4$ & 71 & 2.959 & 18.716 & 2.966 &   -0.450 & 0 & MnB, Ir &FM& 2.003  \\ \hline

Ni$_4$AuB$_4$ & 71 & 3.012 & 18.793 & 2.950 &   -0.224 & 0 & Au, Ni$_4$B$_3$, &NM & 0  \\ \hline

Ni$_4$CuB$_4$ & 71 & 2.992 & 18.125 &2.875&   -0.227 & 0 & B, Cu, Ni$_4$B$_3$ &NM& 0  \\ \hline

Ni$_4$PdB$_4$ &71  & 2.996 & 18.453 & 2.931 &   -0.265 &0  &   Ni$_4$B$_3$, BPd$_2$, B &NM& 0 \\ \hline

Ni$_4$PtB$_4$ & 71 & 2.995 &18.351 &2.960  &   -0.267 & 0 & BPt$_2$, Ni$_4$B$_3$, B  &NM&  0\\ \hline

Ni$_4$ZnB$_4$ & 71 & 2.992 &18.517  & 2.880 &   -0.261 & 0 & Ni$_4$B$_3$, B, ZnNi$_3$B$_2$ &NM&  0 \\ \hline

Fe$_3$Al$_2$B$_2$ $^*$ & 10 &5.685  & 2.833 &8.593  &  -0.426 & 0 &FeAl$_6$, AlB$_2$, FeB  &FM&  0.784  \\
Ref. Exp.~\cite{hirt2018synthesis} &  & 5.723 & 2.857 &2.857&   &  &  &  & & \\ \hline

Fe$_4$Al$_3$B$_2$ & 123 & 8.083 &8.083  &2.791  &  -0.411 & 0 & AlFe, AlFe$_2$B$_2$ &NM& 0.002 \\ \hline

Co$_4$Be$_3$B$_2$ & 123 &7.586  &7.586  & 2.586 &   -0.395 & 0 &Be$_3$Co, BeCo, CoB  &NM& 0  \\ \hline
Ni$_4$Li$_3$B$_2$ & 123 & 8.049 & 8.049 &2.734&   -0.252 & 0 & Li, Ni$_2$B &NM& 0.0002 \\ \hline

Fe$_5$BeB$_2$ & 140 &5.455  & 5.455 &  9.914&   -0.292 &0  &  Be$_2$Fe, Fe$_2$B, Fe &FM& 1.932  \\ \hline

Fe$_5$PB$_2$ $^*$ & 140 & 5.570 & 5.570 & 10.436 &   -0.392 &0.033  & Fe$_2$B, FeB, Fe$_2$P   &FM& 1.705 \\
Ref. Exp.~\cite{haggstrom1975mossbauer}  &  &5.548  & 5.548 & 10.332   &   &  &    &FM&1.730  \\
Ref. Exp.~\cite{mcguire2015magnetic}        &  &5.487  & 5.487 & 10.353   &   &  &   &FM &1.600  \\
Ref. Exp.~\cite{lamichhane2016study} &  &5.485  & 5.485 & 10.348   &   &  &   &FM&1.720  \\
Ref. Exp.~\cite{hedlund2017magnetic}        &  &5.503  & 5.503 & 10.347   &   &  &  &    \\
Ref. Exp.~\cite{cedervall2018influence} &  &5.492  &5.492  &  10.365  &   &  &    &FM&1.658  \\
Ref. Cal.~\cite{werwinski2018magnetocrystalline} &  &5.456  &5.456  &  10.296  &   &  &    &FM&1.770  \\
\hline

Fe$_5$SiB$_2$ $^*$ & 140 & 5.509 & 5.509 & 10.299 &   -0.359 &0.003  & Fe$_2$B, FeSi &FM  &1.731  \\
Ref. Exp.~\cite{mcguire2015magnetic}        &  &5.551  & 5.551 & 10.336   &   &  &  &FM  &1.808  \\
Ref. Exp.~\cite{cedervall2016magnetostructural}        &  &5.554  & 5.554 & 10.343      &  &  & &FM &1.750  \\
Ref. Cal.~\cite{werwinski2016magnetic}       &  &5.546  &5.546  &  10.341  &   &  &  &FM & 1.840 \\ \hline

Co$_5$PB$_2$ $^*$ & 140&5.279  & 5.279 & 10.477 &   -0.357 &0.079  &Co$_2$P, CoB, Co  &FM & 0.409 \\
Ref. Exp.~\cite{rundqvist1962x}        &  &5.420  &5.420  &  10.200  &   &  &  &  &  \\
Ref. Cal.~\cite{werwinski2018magnetocrystalline}       &  &5.284  &5.284  &  10.541 &   &  & &FM &   0.440 \\ \hline

Co$_5$SiB$_2$ $^*$ & 140&5.484  &5.484  &9.942  &   -0.337 &0.042  &CoB, Co$_2$Si, Co   &FM& 0.394 \\
Ref. Exp.~\cite{bormio2010magnetization}        &  &  &  &    &   &  &  & &   \\
Ref. Cal.~\cite{werwinski2016magnetic}       &  &5.511  &5.511  &  9.953  &   &  &  &FM  &0.484  \\ \hline

Cr$_5$PB$_2$ $^*$& 140&5.537  & 5.537 & 10.317 &   -0.474 & 0.032 &Cr$_3$P, CrB  &NM& 0.022  \\
Ref. Exp.~\cite{baurecht1971rontgenographische} &  & 5.593 & 5.593 &10.370&   &  &  & & \\ \hline

Cr$_5$B$_3$ $^*$& 140& 5.431 &5.431  &9.923  &   -0.418 & 0 &CrB, Cr$_2$B  &NM&  0  \\
Ref. Exp.~\cite{guy1976chromium} &  & 5.460 & 5.460 &10.460&   &  &  &  & \\ \hline

Mn$_5$PB$_2$ $^*$& 140 & 5.509 &5.509  &10.287  &   -0.480 & 0.033  &Mn$_2$B, MnB, Mn$_2$P  &FM&  1.665 \\ 
Ref. Exp.~\cite{rundqvist1962x}        &  & 5.540  & 5.540 &  10.490  &   &  &  & &  \\  
Ref. Exp.~\cite{xie2010reversible}        &  & 5.540  & 5.540 &  10.490  &   &  &  & &   \\  \hline

Mn$_5$SiB$_2$ $^*$& 140 & 5.559 & 5.559 & 10.293 &   -0.415 & 0.003 & MnSi, Mn$_2$B &FM&1.583  \\ 
Ref. Exp.~\cite{xie2010reversible}        &  &5.540  & 5.540 & 10.490   &   &  &  &  & \\ 
\hline
\hline
\end{tabular}}
\end{table}

Interestingly, the distance to the convex hull for the experimentally synthesized compounds are finite (Table.~\ref{tab:stable}), {\it e.g.}, Cr$_4$AlB$_6$, Fe$_5$PB$_2$ and Co$_5$PB$_2$ with distances to the convex hull of 12, 33 and 79 meV/atom, respectively. 
This suggests that a loose tolerance of $\Delta E_h < 100$\,meV/atom is more reasonable, taking into account the temperature effects and the systematic errors in DFT calculations.

Critical tolerance with comparable values for the convex hull has also been adopted in other HTP studies~\cite{singh2018high,opahle2013high,wang2017oxysulfide}.
This leads to 434 (335 are MAB phases) stable compounds, as listed in Table S1 in the supplementary. 
As a consequence, our predictions become consistent with another HTP study focusing on Al-containing MAB phases with early transition metals on the \textbf{M}-sites.
For instance, 8 novel MAB phases they found, {\it i.e.}, CrAlB, MnAlB, Cr$_3$Al$_2$B$_2$, Mn$_3$Al$_2$B$_2$, Ni$_3$Al$_2$B$_2$, Mn$_3$AlB$_4$, and Fe$_3$AlB$_4$, are also predicted to be stable using the loose tolerance on the convex hull, as listed in the Table S1. 
It is noted that even if such compounds are metastable, they can still be synthesized using non-equilibrium methods such as MBE and ball milling. 
Hereafter we will consider the stability trend and magnetic properties for all those compounds.
Last but not least, it is essential to consider the non-MAB phases as competing phases beyond those in the OQMD database.
It is observed that the 322-MAB Fe$_3$Al$_2$B$_2$ is stable with $\Delta E_h =0 $ compared with competing phases in OQMD, whereas it becomes metastable with $\Delta E_h = 33 $ meV/atom after considering the non-MAB Fe$_3$Al$_2$B$_2$.
Certainly there are other competing phases and even novel crystal structures beyond those considered in this work, which will be saved for future investigation after experimental validations.

Another interesting question for predicting stable magnetic materials is whether the magnetic configurations
would influence the thermodynamic stability, since most HTP calculations are done assuming the ferromagnetic (FM) 
state as in the OQMD and the Materials Project~\cite{Jain2013}.
This applies particularly to Mn-based compounds, as revealed by a recent work that the energy landscape of the 
convex hull is drastically changed after considering the magnetic ground state~\cite{opahle2020effect}.
According to the literature, the 212-type Mn$_2$AlB$_2$ is observed to display an AFM magnetic ground state
with N\'eel temperature about 390\,K,~\cite{potashnikov2019magnetic, cedervall2019magnetic,ke2017electronic} 
thus we performed extensive calculations on the predicted 212-type MAB compounds.
As summarized in Table.\,S2, 15 out of 54 magnetic compounds prefer AFM magnetic configurations, including not only Mn-based but also Fe- and Co-based compounds.
The magnetic ground states are consistent with those obtained from our Monte Carlo modeling based on the Heisenberg model
taking exchange parameters from DFT calculations (not shown), which will be discussed in detail elsewhere.
Nevertheless, the energy difference between the FM and AFM states is less than 20 meV per atom, 
hence the magnetic ground state has no strong impact on the thermodynamic stability for such compounds.
This can be attributed to the nano-laminated crystal structure, where the magnetic interaction between the local Mn-moments is relatively weak, in comparison to the strongly frustrated fcc-lattice from the Cu$_3$Au lattice considered in Ref~\cite{opahle2020effect}.
It is noted that systematic evaluation of the magnetic ground states is a challenge, hereafter we will focus on the physical properties of the FM states, which should be valid for most of the other compounds.

After the thermodynamic stability, mechanical and dynamical stabilities should be addressed as well in order to make systematic predictions.
It is observed that mechanical stability plays a marginal role as explicitly demonstrated for 21 stable compounds on the convex hull.
This is consistent with our previous studies on the antiperovskite compounds.~\cite{singh2018high}.
For the orthorhombic MAB phases, there are nine independent elastic constants C$_{11}$, C$_{22}$, C$_{33}$, C$_{44}$, C$_{55}$, C$_{66}$, C$_{12}$, C$_{13}$, and C$_{23}$. For the tetragonal non-MAB phases, there are six independent elastic constants C$_{11}$, C$_{33}$, C$_{44}$, C$_{66}$, C$_{12}$, and C$_{13}$.
According to the  mechanical stability defined in the Ref.~\cite{mouhat2014necessary}, none of the novel compounds predicted to be thermodynamically stable is found to be mechanically unstable.
In addition, the dynamical stability is verified by examining the the phonon spectra as compiled in Fig. S2  for 21 predicted and 15 known cases.
Obviously, there is no imaginary modes observed for 35 compounds, indicating that those compounds are stable against local atomic displacements.
The resulting phonon spectra for Cr$_2$AlB$_2$ and Cr$_3$AlB$_4$ are in good agreement with previous reported results~\cite{bai2019phase}.

Nevertheless, for Ni$_4$Li$_3$B$_2$ there exists an imaginary mode at the M point. This suggests that the compound may end up with other crystal structures or synthesized on certain substrates using molecular beam epitaxy.

\subsubsection{Trends in the stability}

To understand the trend of stabilities for the MAB and non-MAB phases, the number of stable compounds ($\Delta E_h < 100$  meV/atom) with respect to the \textbf{A} element are shown in Fig. S4.
It is obvious that most elements in the periodic table act as a constituent element stabilizing at least one of the considered crystal structures, whereas nine out of 59 elements ({\it i.e.}, K, Rb, Cs, Sr, Ba, Zr, N, Sb, and Bi) do not form any stable phases.
Particularly, each of the five elements like Be, Al, Pt, Zn, and Ir support more than 22 stable phases.
Moreover, among all the structure types considered in this work, 136 compounds are stable with the 414-type structure, albeit the first compound Cr$_4$AlB$_4$ was reported in 2019.~\cite{zhang2019crystal}.

\begin{figure}
    \centering
    \includegraphics[width=15cm]{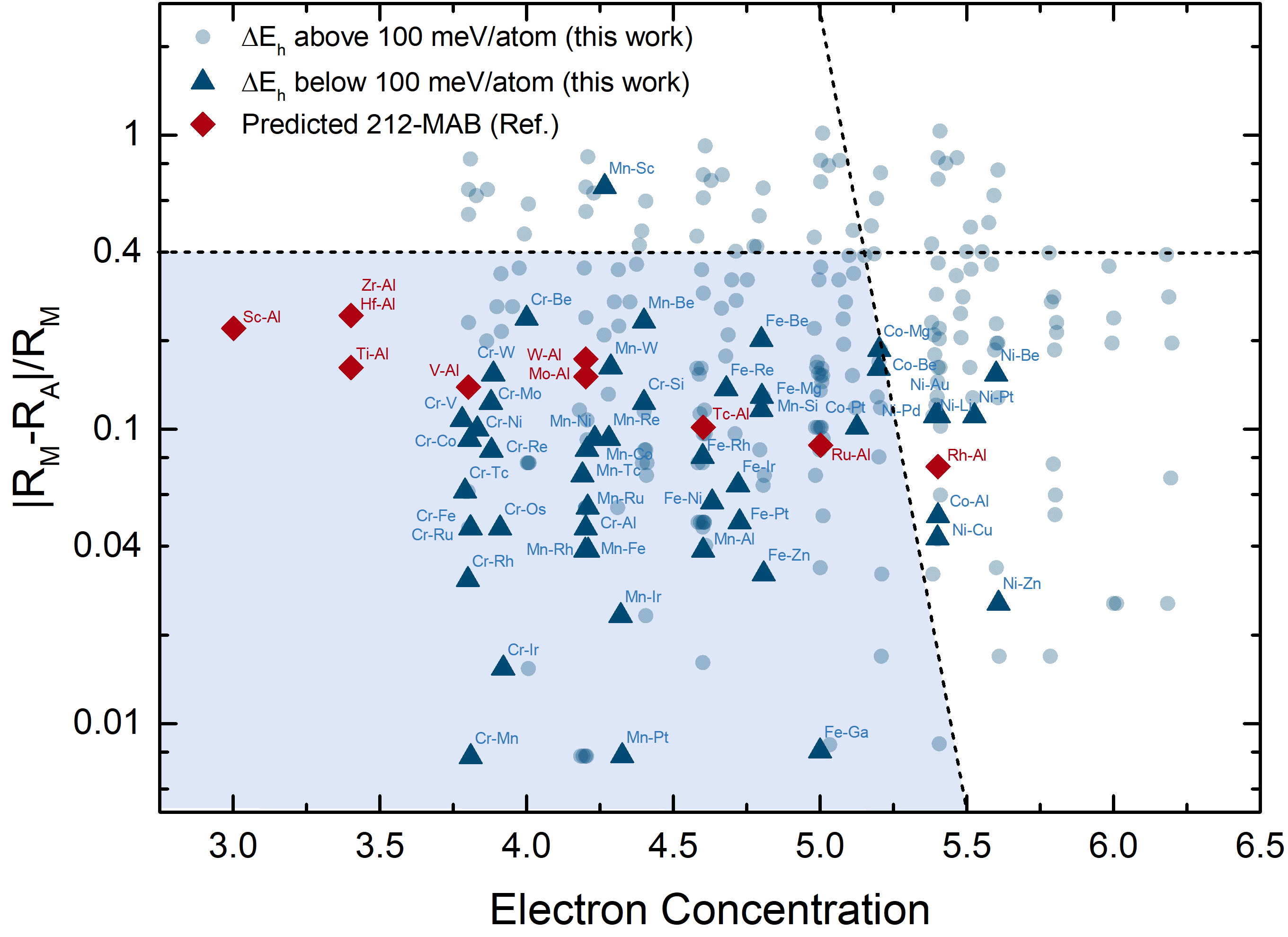}
    \caption{The stability map of 212-MAB phases (circle symbols represent unstable phases in the present work; triangle symbols represent possible stable phases with convex hull distance below 100 mev/atom in the present work; square symbols represent newly reported novel phases in Ref.~\cite{khazaei2019novel}).
    }
    \label{formability}
\end{figure}

Taking the 212-type MAB structure as an example, the stability trend with respect to the chemical composition can be understood based on the Hume-Rothery rules~\cite{mizutani2012hume}.
Such rules are formulated based on the difference of size, electronegativity factors and the valence electron concentration (VEC).
It is observed that the electronegativity difference between the M and A elements has no strong correlation with the stability (Fig. S5), same as the MAX compounds.~\cite{zhang2020role}
On the other hand, as shown in Fig.~\ref{formability}, both the atomic radius difference of the M and A elements $\frac{|R_\text{M}-R_\text{A}|}{R_\text{A}}$ and VEC have significant influence on the stability.
Clearly, most stable compounds are within the region $\frac{|R_\text{M}-R_\text{A}|}{R_\text{A}} < 0.4$ and VEC $<$ 5.5.
The newly reported novel phases in Ref.~\cite{khazaei2019novel} also prove the practicality of the current expression factors.
Similar behavior is also observed for the 414-type MAB compounds with a slightly smaller tolerance for VEC $<$ 6, as shown in Fig. S6.
The reason might be due to the fact that the \textbf{M}-site and boron-site contributing less valence electrons because of the extended M-B block (Fig.~\ref{all-structure}).

The general trend in the stability can be
elucidated based on the chemical orbital Hamilton population (COHP) obtained using the LOBSTER code,~\cite{maintz2016lobster} which provides an atomic picture about the bonding.
As an example which is representative for all the compounds we considered, the bond-resovled COHP is shown for M$_2$AlB$_2$ (where M are Cr, Mn, Fe, Co and Ni) and Fe$_2$AB$_2$ (where A are Be, Mg, Ca, Sr and Ba) in Fig. S8. 
Focusing on varying the \textbf{M} elements, the number of valence electron on the \textbf{M}-sites increases from 6 in Cr$_2$AlB$_2$, to 8 in Fe$_2$AlB$_2$, and finally to 10 in Ni$_2$AlB$_2$. 
For Cr$_2$AlB$_2$, it is obvious that the values of -COHP are all positive below the Fermi energy, indicating only boning states are occupied in the corresponding bonds, which leads to a high overall stability (Fig.~S8)
Increasing the number of valence electron to 10 in Ni$_2$AlB$_2$, the negative energies -COHP appeared below the Fermi energy in the Ni-B, Ni-Al, and Ni-Ni bonds. 
The occupation of such anti-bonding states weakens the bonds and therefore destabilizes the Ni$_2$AlB$_2$ compound.
Therefore, the ICOHP of M-Al and M-B are increasing within the the number of valence electron increasing, which indicates the corresponding bond strength weakens.
Similar behaviour is also observed in the COPH os Fe$_2$AB$_2$ compounds with varying A elements being Be, Ca, and Ba (Fig.~S8).
As the atomic size changes from 0.99 $\AA$ (Be), 1.74 $\AA$ (Ca) and 2.06 $\AA$ (Ba), the bond strength of those compounds becomes weaker, which are confirmed by the COHP values of Fe-Fe, Fe-B, Fe-A and A-B.

Hence, with respect to varying both M and A elements with increasing number of valence electrons and atomic size, the Fermi energy is shifted into the anti-bonding states, leading to instability.
This helps to understand the trend observed in Fig.~\ref{formability}, which provide valuable guidance to guide the synthesis of MAB phases by substituting the M/A sites or via forming solid solutions.

\subsection{Magnetic properties}

\subsubsection{MAE}
\begin{figure}[bh!]
    \centering
    \includegraphics[width=15cm]{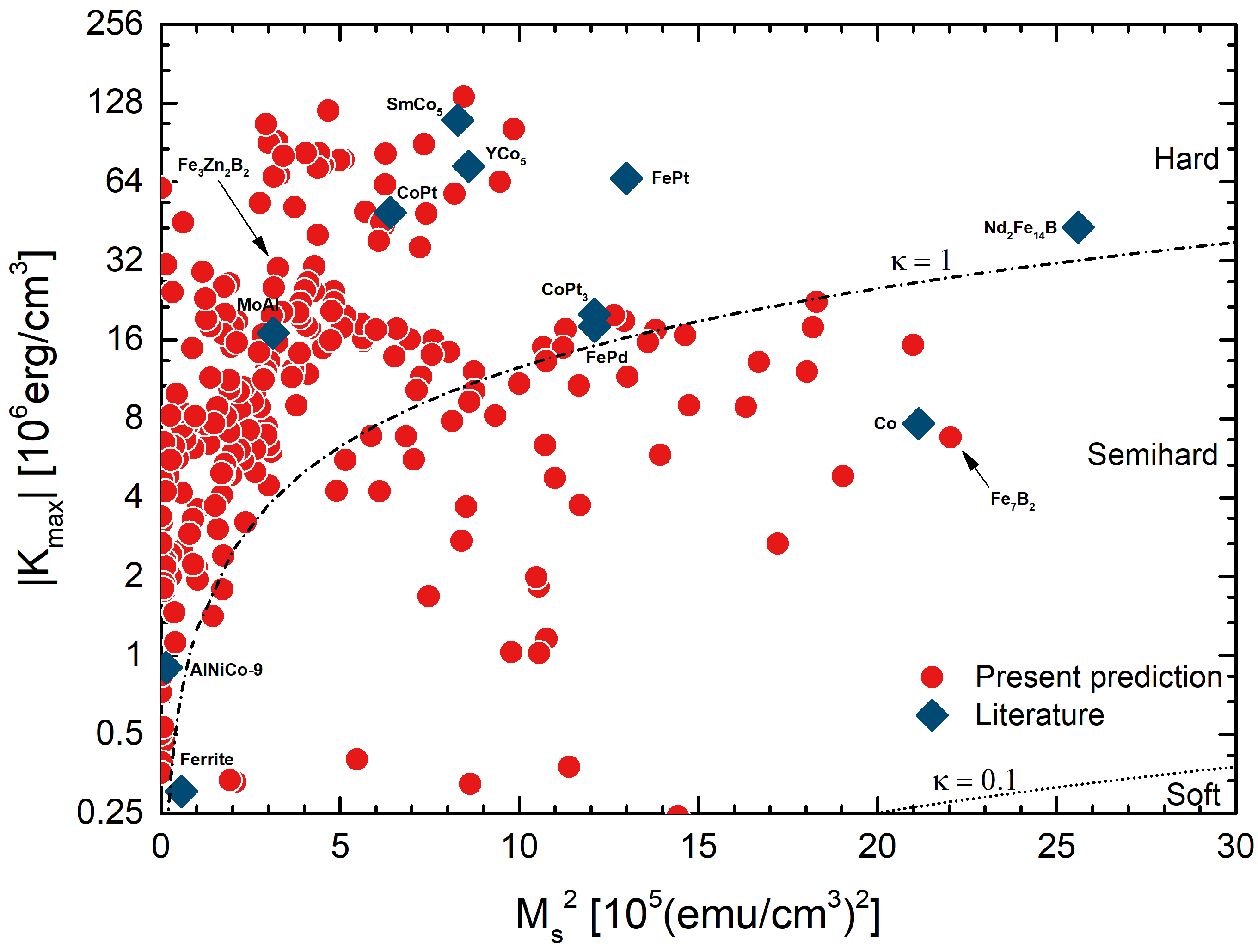}
    \caption{The MAE vs magnetization of the promising candidates of targeted phases. The dashed lines correspond to magnetic hardness parameter $\kappa = \sqrt{K_1/(\mu_0M_S^2)}$ for values $\kappa$ = 1 and 0.1. Hard magnetic materials ($\kappa>1$) can be used to make efficient permanent magnets of any shape.
    }
    \label{maevsms}
\end{figure}

Turning now to the magnetic properties, we focus on the magnetocrystalline anisotropy energy (MAE) and magnetocaloric effect (MCE), in order to identify potential candidates for permanent magnet and magnetocaloric applications.
The MAE is caused by the broken continuous symmetry of magnetization directions due to the spin-orbit coupling (SOC),~\cite{bruno1989tight} 
which is defined (denoted as K) in terms of 
\begin{equation}
    \text{K}_{\hat n_1-\hat n_2}= \text{E}_{\hat n_1}-\text{E}_{\hat n_2},
\end{equation}
where $\text{E}_{\hat n}$ denotes the total energy with the magnetization direction parallel to $\hat n$. In the present work, we consider $\hat n$ along three crystalline directions, namely, [100], [010] and [001], as MAB compounds have orthorhombic structures (Fig.~\ref{all-structure}).
This leads to three MAEs, {\it i.e.}, K$_{001-010}$, K$_{001-100}$ and K$_{010-100}$. 
Fig.~\ref{maevsms} shows the MAE with respect to the saturation magnetization (M$_\text{S}$), in comparison with the experimentally known permanent magnets.
There are in total 71 cases (cf. Table. S6 in the supplementary) with the absolute value of at least one MAE greater than 1.0 MJ/m$^3$.
For instance, the MAE of Fe$_2$AlB$_2$ has been evaluated by different groups~\cite{ke2017electronic, cedervall2016magnetic, barua2018anisotropic}, and our result of -1.14 MJ/m$^3$ is in good agreement with the experimental measurements of -0.9 MJ/m$^3$ at 50 K by Barua~\cite{barua2018anisotropic} and theoretical calculation -1.34 MJ/m$^3$ by Ke~\cite{ke2017electronic}.
In the newly predicted compounds, the MAB phase Mn$_4$PtB$_4$ has the largest MAE as 13.498, 11.948 and -1.550 MJ/m$^3$ for K$_{001-010}$, K$_{001-100}$ and K$_{010-100}$.
Additionally, the 111-type FePtB shows the largest MAE in non-Mn-containing compounds as -10.646, 7.225 and -3.421 MJ/m$^3$ for K$_{010-100}$, K$_{001-010}$ and K$_{001-100}$, suggesting the \textit{b}-direction (c-direction) is easy (hard) axis.
Based on the dimensionless figure of merit $\kappa = \sqrt{K_1/(\mu_0M_S^2)}$~\cite{coey2014new}, 
there exist quite a few compounds which can be classified as hard magnets.
Particularly, the MAE of such ternary TM borides fill the gap between the widely used low performance magnets (such as AlNiCo and ferrite) and high performance magnets (such as Sm-Co and Nd-Fe-B). 

However, not only the absolute values of the MAE but also the sign matters, {\it e.g.}, the easy axis (direction with the lowest energy) is ideally aligned along a special crystalline axis.
For all the MAB compounds, the [001] direction along the stacking direction of the M-B layers (Fig.~\ref{all-structure}) is chosen, corresponding to the most-probably exposed surfaces for such nano-laminated structures. 
For the non-MAB phases of the tetragonal space groups, the special axis is chosen to be the axis of 4-fold rotational symmetry, {\it i.e.}, the [001] direction in Fig.~\ref{all-structure}(g \& i).
The MAE for the 322-type compounds (Fig.~\ref{all-structure}(h)) is overall small thus we do not consider them. 
Correspondingly, we found 16 MAB and 7 non-MAB phases with a significant out-of-plane MAE ($>$ 0.4 MJ/m$^3$),
as well as 33 (18) MAB (non-MAB) compounds with a reasonable in-plane MAE (absolute value larger than 0.4 MJ/m$^3$), as listed in Table S7. 
Among them, the 322-type MAB compound Mn$_3$Ir$_2$B$_2$ has the largest out-of-plane MAE of 10.17 MJ/m$^3$ for K$_{010-001}$, and Fe$_2$ReB$_2$ with a large MAE of 9.00 MJ/m$^3$ in K$_{010-001}$. 
Interestingly, the MAE value of Fe$_3$Zn$_2$B$_2$ is as large as 3.00 MJ/m$^3$ in K$_{100-001}$ while its M$_s$ is comparable to that of MnAl.
It contains no expensive, toxic or critical element, which is a good candidate permanent magnet material.
Moreover, Fe$_7$B$_2$ has a sizable MAE 0.681 MJ/m$^3$, which is quite comparable to that of hcp Co. 
Such a phase is beyond the known binary Fe-B phase diagram~\cite{hallemans1994thermodynamic}, which might be synthesizable under non-equilibrium conditions.
Last but not least, the diagram (Fig.~\ref{maevsms}) together with Table S6 give us a chart to engineer permanent magnets via doping. 
For instance, our calculations reveal that Fe$_5$PB$_2$ has an MAE of 0.63 MJ/m$^3$ consistent with the experimental measured value of 0.65 MJ/m$^3$~\cite{hedlund2017magnetic}, whereas a recent work demonstrated that its MAE can be enhanced by substitutionally doping tungsten~\cite{thakur2019enhancement}. 

As discussed above, most compounds with significant MAE contain $5d$ elements, such as Pt, Ir, and Re. 
This suggests that the MAE is originated from the enhanced atomic SOC strength for the $5d$-shell of such elements.
Following Ref.~\cite{antropov2014constituents}, the atomic resolved SOC energy changes are listed in Table. S8 for the 111-type FeXB with X = Ni, Pd, and Pt.
As the atomic SOC strength increases from 98 meV for Ni, 185 meV for Pd, to 533 meV for Pt~\cite{richter1998band}, 
the contribution from the X element to the MAE is becoming more significant, as given by the change of atom-resolved SOC energy $\Delta \text{E}_\text{SOC}=\text{E}_\text{SOC}(\hat n_1) - \text{E}_\text{SOC}(\hat n_2)$.
\textcolor{black}{For FeNiB, $\Delta$E$_{SOC}(\text{Fe})$ (-0.492 meV/at. in [100]-[010] direction) dominants the total $\Delta$E$_{SOC}$ (-0.586 meV/f.u. in [100]-[010] direction) of the compound, as the SOC strength is comparable for Fe (55 meV) and Ni.
Furthermore, for FeXB with X = Ni, Pd, and Pt, the $\Delta$E$_\text{SOC}$ of Ni, Pd, and Pt are -0.093, 0.702, and 2.603 meV/atom between two magnetization directions [100] and [010], corresponding to the changes in the total MAE of -0.128, 0.181, and 2.106 meV/atom, respectively.
That is, $\Delta$E$_{SOC}$ of X has a more dominant contribution to the total $\Delta$E$_\text{SOC}$ and hence the MAE, when moving down the group from $3d$ to $5d$ elements. 
In the FePtB, the contribution of $\Delta$E$_{SOC}$ of Pt is 84\% in total $\Delta$E$_{SOC}$.
Therefore, like FePt~\cite{daalderop1991magnetocrystalline}, the $5d$ elements have a more significant contribution to the MAE because of enhanced atomic SOC strength, though the magnetic moments on such elements are induced by those of the $3d$ atoms.}

\subsubsection{MCE}
As introduced above, it is postulated that ternary TM borides are promising candidates for MCE applications, such as Fe$_5$SiB$_2$~\cite{cedervall2016magnetostructural} and Fe$_2$AlB$_2$~\cite{tan2013magnetocaloric,barua2018anisotropic,el2019magnetocaloric,cedervall2016magnetic}.
To search for more candidates in the predicted MAB and non-MAB compounds, we performed screening based on the magnetic deformation proxy~\cite{bocarsly2017simple}.
It is demonstrated that the magnetic entropy change $\Delta S_M$ upon magneto-structural transitions has a strong correlation with the magnetic deformation $\Sigma_M=\frac{1}{3}(\eta_1^2+\eta_2^2+\eta_3^2)^{1/2}\times100$ and $\boldsymbol{\eta}=\frac{1}{2}(\mathbf{P}^T\mathbf{P}-\mathbf{I})$ 
where $\mathbf{P}=\mathbf{A}_\text{nonmag}^{-1}\cdot\mathbf{A}_\text{mag}$ 
with $\mathbf{A}_\text{nonmag}$ and $\mathbf{A}_\text{mag}$ being the lattice constants of the nonmagnetic and magnetic unit cells.
Although there is no direct scaling between $\Delta S$ and $\Sigma_M$, it is suggested that $\Sigma_M > 1.5\%$ is a reasonable cutoff to select the promising compounds.~\cite{bocarsly2017simple}

\begin{figure}
    \centering
    \includegraphics[width=15cm]{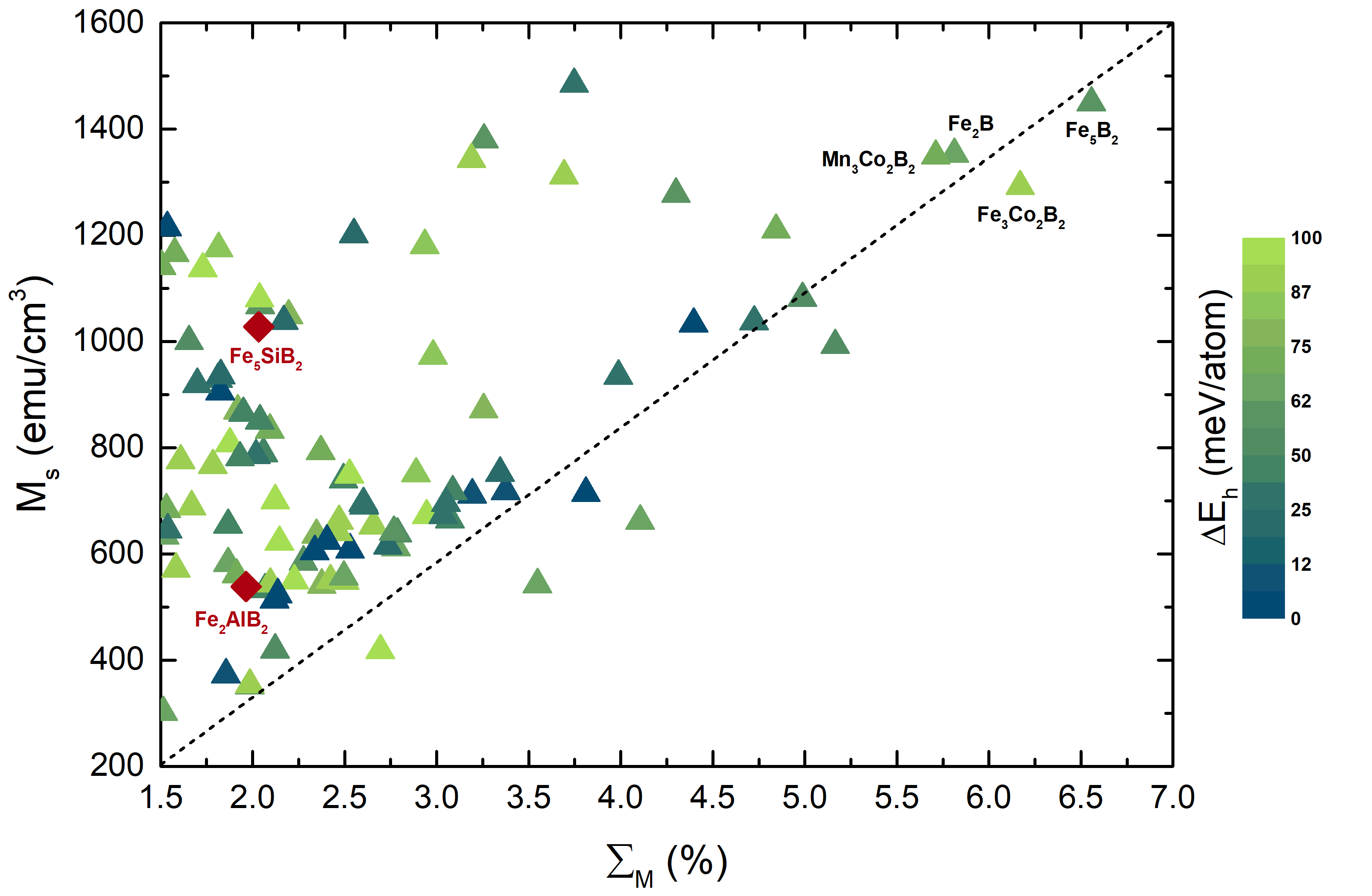}
    \caption{The 99 potential MCMs with magnetic deformation $\sum_M > 1.5 \%$. The color bar marks the distance to the convex hull. The dash line indicates a positive correlation between the magnetization density and the magnitude of magnetic deformation.
    }
    \label{Ms_deform}
\end{figure}

Fig. \ref{Ms_deform} shows the 99 potential MCMs with $\sum_M > 1.5 \%$ from 434 compounds with convex hull $\Delta E < 100$ meV/atom. 
Among them, the reported~\cite{bocarsly2017simple} $\sum_M$ of Fe$_5$SiB$_2$ (2.14\%) and Fe$_2$AlB$_2$ (2.05\%) are confirmed in our calculations, with the resulting $\sum_M$ of 2.03 \% and 1.96 \%, respectively. 
Interestingly, there is positive correlation between the magnetization density and the magnitude of magnetic deformation, {\it i.e.},  as the magnetic deformation increases, the magnetization of compounds also increase (Fig.~\ref{Ms_deform}). 
It is noted that 82 out of 99 potential MCMs locating at $\sum_M < 3.5\%$, and the magnetization concentrating between 500 to 1000 emu/cm$^3$. 
Particularly, there are four compounds, {\it e.g.}, Fe$_5$B$_2$ (322-MAB), Fe$_3$Co$_2$B$_2$ (322-MAB), Mn$_3$Co$_2$B$_2$ (322-MAB), and Fe$_2$B (111-MAB),  at the upper-right corner, which perform on both large magnetization and  magnetic deformation. 
We suspect such compounds can exhibit significant $\Delta S_M$ upon second order phase transition at the corresponding Curie temperature, which will be saved for detailed investigation in the future.
\\
Several important aspects on possible MCE in such materials are noteworthy, based on the distributing map with respect to the  \textbf{M} and \textbf{A} sites as shown in Fig. S9. 
For instance, compounds with Fe and Mn occupying the \textbf{M}-site show a high possibility to posses a large MCE based on the magnetic deformation, which have been confirmed in several reported compounds~\cite{xie2010reversible,arora2007magnetocaloric,yu2003large}.
Based on the correlations observed in known MCMs in Ref.~\cite{bocarsly2017simple}, such materials are likely to show a strong magnetocaloric effect and are therefore excellent candidates for experimental study.
Moreover, compounds with Mn/Fe/Co, Ru/Rh/Pd and Os/Ir/Pt occupying the \textbf{A}-site also show a high potential to host remarkable magnetocaloric properties.
Furthermore, it is noted the fact that Fe$_2$AlB$_2$ is composed entirely of earth-abundant elements.
This provides a major advantage at least from a cost and resource point of view, over the competing MCMs that contain expensive critical elements ({\it e.g.}, Gd, Gd$_5$Si$_2$Ge$_2$, FeRh).
Therefore, such economic material without critical elements appears especially appealing to us, and the present system M$_x$A$_y$B$_z$, when A = Al, Zn, Si and Fe should be attracted more attention,
such as Fe$_4$AlB$_4$ (2.33 \%), Fe$_3$AlB$_4$ (2.11 \%), Fe$_4$SiB$_4$ (2.73 \%),  Fe$_3$ZnB$_4$ (2.42 \%) and Fe$_5$B$_2$ (\textit{Cmmm}, 6.56 \%) (Fig. S9).

\section{Conclusion}
In summary, our high-throughput screening on 6 types of MAB phases and 3 types of competing non-MAB phases predict 434 magnetic ternary transition metal borides which are potential candidates for permanent magnets and magnetocaloric materials.
After validating the 15 reported compounds, 21 novel compounds are identified to be stable based on the systematic evaluation of thermodynamic, mechanical, and dynamical stabilities, 
and the number of stable compounds is increased to 434 taking the tolerance of convex hull being 100 meV/atom.
It is observed that the magnetic ground state for such compounds with layered structures does not have a strong influence on the thermodynamic stability.
The trend of stability for the MAB phase can be understood based on the Hume-Rothery rules, where the size factor difference and the valence electron concentration play a critical role.
Such a trend can be further attributed to the bond-resolved COHP, providing intuitive guidance for experimental synthesis.
The detailed evaluation of the magnetocrystalline anisotropy energy and the magnetic deformations leads to 23 compounds with significant uniaxial anisotropy ($>$ 0.4 MJ/m$^3$) and 99 systems with reasonable magnetic deformation ($\sum_M > 1.5 \%$).
For those compounds containing no expensive, toxic, or critical elements, it is observed that Fe$_3$Zn$_2$B$_2$ is a reasonable candidate as gap permanent magnet, and Fe$_4$AlB$_4$, Fe$_3$AlB$_4$, Fe$_3$ZnB$_4$, and Fe$_5$B$_2$ as potential magnetocaloric materials. 
This work paves the way for designing more magnetic materials for energy applications and magnetic MBene for two-dimensional magnetic materials.
At last, the realistic assessment of the predicted potential MAB phases are conducting and will add to our library~\cite{kuz2014towards,gottschall2019making} soon.

\begin{acknowledgement}
The authors gratefully thank Manuel Richter for providing the SOC strength data and helpful discussions.
This work was also supported by the Deutsche Forschungsgemeinschaft (DFG, German Research Foundation) - Project-ID 405553726 - TRR 270.
Part of this work was supported by the European Research Council (ERC) under the European Union’s Horizon 2020 research and innovation program (Grant No. 743116-project Cool Innov)
The Lichtenberg high performance computer of the TU Darmstadt is gratefully acknowledged for the computational resources where the calculations were conducted for this project.

\end{acknowledgement}

\begin{suppinfo}
Numerical details, list of all metastable compounds, and additional theoretical results can be found in the Supplementary Material.
\end{suppinfo}

\bibliography{achemso-demo}

\end{document}